# Development and validation of electrical-insulating $Al_2O_3$ coatings for high-temperature liquid PbLi applications

Abhishek Saraswat[1,3], Chandrasekhar Sasmal[1], Ashokkumar Prajapati[1], Rajendraprasad Bhattacharyay[1], Paritosh Chaudhuri[1,2] and Sateesh Gedupudi[3]

[1] Institute for Plasma Research, Gandhinagar-382428, India

[2] Homi Bhabha National Institute, Trombay, Mumbai-400094, India

[3] Heat Transfer and Thermal Power Laboratory, Department of Mechanical Engineering,

Indian Institute of Technology Madras, Chennai-600036, India

asaraswat@ipr.res.in

**Abstract.** Electrical-insulating coatings are of great importance for applications in liquid-metal breeder/coolant based systems relevant to nuclear fusion power plants. In specific to lead-lithium eutectic (Pb-16Li), a candidate breeder material, such coatings are being actively investigated worldwide for their criticality in addressing various functionalities including, but not limited to, reduction in Magneto-Hydro Dynamics (MHD) pressure drop, corrosion resistance to structural materials and development of specific diagnostics, like two-phase detection techniques. For such applications, a candidate coating must be demonstrated for its compatibility with corrosive media, endurance towards high operational temperatures and integrity of electrical-insulation over long operational durations without substantial degradation. At present, no relevant in-situ insulation resistance data is available for quantitative performance assessment of coated substrates within PbLi environment over long operational durations. To address this shortfall, an experimental study was performed at Institute for Plasma research (IPR) towards application of high-purity alumina ($Al_2O_3$) coatings on SS-316L substrates and further rigorous validation in static PbLi environment. The adopted coating process required a low-temperature heat treatment ($< 430°C$) and could yield average coating thicknesses in the range of $\sim 100\ \mu m - 500\ \mu m$. Coated samples were validated for their electrical insulation integrity in static PbLi over two test campaigns for continuous durations of over 700 h and 1360 h, including thermal cycling, at operational temperature in the range of $300°C$-$400°C$. Volumetric electrical-resistivity, estimated through high-voltage insulation resistance measurements at relevant temperatures, remained of the order of $10^9$-$10^{11}$ $\Omega$-cm without significant degradation. In-situ estimations of thermal derating factors establish excellent electrical-insulation characteristics after long term exposure to liquid PbLi. This paper presents details of utilized coating application methods, coating thickness estimations, liquid-metal test set-up, insulation performance and critical observations from SEM/EDX and XRD analysis on the tested samples.

## 1. Introduction

Liquid-Metal (LM) based breeding blanket concepts like HCLL (Helium-cooled Lithium Lead), WCLL (Water-cooled Lithium Lead), DCLL (Dual Coolant Lithium Lead) and LLCB (Lead-lithium Ceramic Breeder) provide several key advantages including low vapor pressure, high thermal conductivity and stability under radiation environment typical for a fusion reactor. However, presence of a transverse magnetic-field significantly modifies flow geometry of an electrically-conductive fluid due to imposed Lorentz force, posing high demands on the required pumping power to compensate for resulting pressure drops. Successful materialization and utilization of such LM ancillary systems in a fusion power plant requires optimization of process parameters including MHD pressure drop reduction, which mandates an electrical-isolation between the LM and structural material [1-24]. Such an electrical-insulation, by virtue of its non-porosity towards LM ingress, is also envisaged to act as a corrosion-protection layer as well as Tritium Permeation Barrier (TPB) [4, 8-11, 14, 20, 25-37]. Additionally, development of a few specific diagnostic tools for LM systems necessitates provision of electrical-isolation over a substantial length [38-42].

Numerous experimental and simulation studies are being carried out worldwide towards performance qualification of functional materials which can cater to above needs either in the form of a coating or as Flow-Channel Inserts (FCIs), with $Al_2O_3$, $Er_2O_3$, SiC identified as potential candidates in particular [9-10, 13, 43, 44]. However, a scarcity of relevant experimental data can be observed towards quantitative performance assessment of electrical Insulation Resistance (IR) for coated substrates over long duration exposure to Pb-16Li (hereafter referred to as PbLi) environment at relevant temperatures. IPR is working in the areas of MHD, corrosion studies of structural materials and development of a two-phase detection technique relevant to liquid PbLi applications [45-47]. In this view, to address the crucial requirements of electrical-insulation for further advancements in LM domain, a preliminary experimental study was conducted towards rigorous qualification of $Al_2O_3$ based electrical-insulating coatings deposited over SS-316L substrate electrodes. $Al_2O_3$ was preferentially chosen as the coating material of interest for its thermodynamic and chemical stability while the selection of substrate material was largely dominated by its commercial availability, inexpensiveness and proven usage as structural material and wetted parts of diagnostics in small scale PbLi R&D facilities. One of the major challenges in the application of ceramic coatings over SS-316L substrates is the wide difference between Coefficients of Thermal Expansions (CTEs) of the two materials over operational temperature range of interest [48]. Through various experimental runs, optimization of heat-treatment parameters was achieved successfully to develop $Al_2O_3$ coated SS-316L electrodes. Adopted coating application process allows for desired variations in the yielded coating thicknesses. Details of the application methods, performance tests in static liquid PbLi and metallurgical observations are presented in the following sections.

## 2. Experimental methods and materials

Previous works have well-demonstrated successful $Al_2O_3$ growth over substrates of SS-304L, SS-316L, P-91 and RAFM (Reduced Activation Ferritic-Martensitic) steels alongwith an excellent corrosion resistance observed against static and flowing PbLi [4, 27, 33 49-51]. A few efforts for shorter durations, limited to ~200 h, have also been made towards in-situ validation of solid high-purity $Al_2O_3$ pieces as electrically-insulating FCI in PbLi environment [2]. However, no reported study provides relevant long duration performance data on IR for $Al_2O_3$ coated substrates to address critical issues related to MHD pressure drops and development of diagnostics, as highlighted above. Additionally, during the present study, it was repeatedly observed that the plasma assisted tempering and thermal tempering [49, 52] of hot-dip aluminized SS-316L substrates resulted in an electrically conductive layer, which could be verified even at a macroscopic level, suggesting incomplete transformation to $Al_2O_3$. The study also resulted in substrate deformation due to high-temperature exposure. It may therefore be concluded that such coating application methods demand independent sophisticated optimization of parameters for different substrate materials. However, this further needs to be investigated in detail. In the present study, with an aim to homogenize the coating application method towards various substrate materials, a refractory coating suspension of aluminium monophosphate-bonded calcined $Al_2O_3$ with 99.8% purity (containing ~ 75% solids by weight) has been investigated on 1.6 mm thick solid electrodes of SS-316L.

Pre-coating preparation of the substrate includes surface roughness induction using grade-80 emery paper followed by acetone application for removal of dust, grease and foreign particles to allow for a better adhesion between coating and substrate. The coating was applied in two steps: (i) dip-coating for 30 seconds with gradual retraction and (ii) manual brush-coating on the surface of the electrode. To ensure the experimental objective of in-situ IR measurements in electrically conducting PbLi, tip of the electrode was provided with an additional dip-cycle in the same coating suspension, resulting in a blob-shaped coating deposition. Thereafter, coated samples were provided an air-set for ~ 24 h in a vertical orientation (coated probe-tip at the top). Gradual retraction of the coating material under gravity allows the suspension to fill any remnant voids and undulations intentionally induced by surface roughening. This air-set period, acting as a necessary drying and pre-bonding stage, was followed by high-temperature heat-cure in a muffle furnace with the coated electrode(s) arranged in a horizontal orientation. High-temperature heat-cure involves following stages:

a) Ramp-up time of 2 h from room temperature (RT) to 93°C followed by 2 h heat-cure at 93°C. This step was intended towards gradual dehydration of the coating.

b) Ramp-up time of 2.5/10/12 h from 93°C to 427°C followed by 2 h heat-cure at 427°C. This step was intended towards achieving a good bond between $Al_2O_3$ and SS substrate.

c) A natural cool-down from 427°C to RT.

Post heat-treatments, absence of any voids/cracks was ensured through visual inspections, which was followed by high-voltage IR measurements at RT for one of the fabricated probes. The in-situ IR measurement scheme for PbLi environment is depicted in **Fig.1**. To the best of the authors' knowledge, this is the first time in-situ high-voltage IR measurements are employed for $Al_2O_3$ coated substrates in PbLi environment over long duration experimental studies. Test set-up consists of a PbLi tank provided with surface mount electrical resistive-heater in a temperature feedback control loop and ports for vacuuming, gas feeding/purging and venting. Three coated probes, namely P1, P2 and P3, were fabricated as per the procedure described above. However, for P2 & P3, fully heat-cured layers were further deposited with an additional brush-coated layer to allow gradual heat-curing through system operations over long duration. P2 was installed as-such in the PbLi tank while P3 was first dehydrated at 93ºC for 2 h (step (a)) prior to installation in the tank. A bare SS-316L electrode (A), inserted through the third port, acts as reference conducting electrode for IR measurements. PbLi in the tank was melted and heated upto a temperature of 300°C over a duration of ~15 h in ultra-high purity (UHP) grade argon gas environment with repeated argon purging to maintain an inert environment. A positive cover gas pressure was maintained over complete test-duration. After liquefaction of PbLi (melting point observed at ~235°C-236°C), coated probes and the bare SS-316L electrode (A) were immersed in PbLi. The electrical resistance between A and the test facility structure was < 1 Ω, ensuring a good electrical contact. Leverage of high electrical-conductivity of PbLi [53-54] was taken to measure IR of immersed probes being equal to the measured resistance between accessible conducting parts of the coated probes (Px, x = 1,2,3) and the electrode A. P1 and P2 were tested during test campaign-1 while P3 was tested during campaign-2. High voltage insulation tests were performed at regular intervals over complete test campaigns including thermal cycles. Temperature derating factors, signifying the ratio of IR at any given temperature to the IR value at base temperature, were estimated during last thermal cycle for every 10°C rise in temperature from the base temperature. Test-campaigns were concluded with final IR measurements to observe cumulative degradation effect on the IR due to thermal aging, corrosion and thermal cycling. Post-completion of test-campaigns, samples of ~ 8 mm length were cut across the cross-sections using a diamond wire-cutter (Make: Gatan, wire diameter: 200 µm) followed by gold sputter-coating of the samples for Scanning Electron Microscopy (SEM) and Energy Dispersive X-Ray (SEM-EDX) analysis. Metallurgical analyses on the prepared samples were carried out for coating thickness estimations, microstructure evaluations and LM ingress detection. Particle size measurements were performed using SEM- image analysis software and open source imageJ software tool.

## 3. Results and discussions

### 3.1 Coating observations

The adopted dip-coating method resulted in a relatively thinner (~150-200 µm) wet-film layer over the substrate when the electrodes were retracted very gradually. As depicted in **Fig.2**, a short ramp-up time (2.5 h) for heat-treatment (step (b)) resulted in visible cracks along the coated length with formation of surface blisters, which can be attributed to a larger differential thermal expansion and sudden dehydration of coated suspension, respectively. Although a longer ramp-up time (~ 10 h to 12 h) greatly minimized the cracks and blisters during first heat-cure, elimination of very fine cracks could not be achieved. Hence, additional coating layers deemed necessary repeating the application steps until no visible cracks were identified. The IR at RT for one of the probes, prepared with a single-dip coat cycle followed by full heat-curing, was found to be exceptionally high providing an insulation leakage current of < 1 nA. In view of highly brittle nature of $Al_2O_3$, preliminary non-destructive investigations were carried out using a calibrated portable USB microscope for estimation of coating thicknesses using average diametrical measurements as shown **in Fig.3 (a)** and **Fig.3 (b)**. Detailed average coating thickness estimations across the cross-section (destructive examinations) using SEM for the tested probes are presented under dedicated section on metallographic investigations. Details of the probes tested for compatibility with PbLi are summarized in **Table-1** below. All weight measurements were done using a calibrated precision weighing balance (Sartorius make, CPA225D) with a resolution of 10 µg.

### 3.2 Performance validations in liquid PbLi environment

#### 3.2.1. Test campaign-1

Temperature variations with time duration details for the test campaign-1 are depicted in **Fig.4**. The red-markers (circle) on the plot represent instances of in-situ high-voltage IR tests. Measured IR values with time and temperature are mentioned in **Table-2**. All IR measurements were taken 60 s after high-voltage application to minimize the effects of capacitive charging current and dielectric absorption current [55].

A decrease in IR was observed for both P1 and P2 with an increase in temperature from 300°C to 350°C, an expected temperature-based derating phenomenon in insulators. However for P2, during 260 h in static PbLi at 300°C, an increase in IR value was observed which could be accounted for gradual heat-curing of uppermost untreated coat layer over long operational durations. From the measured performance data, immediate observable advantages of the adopted coating method include selective coating/masking on difficult substrate geometries including the internals of pipe-sections and complex manifolds, easy application without exposure of substrate to very high temperatures as required in previous works [11, 14, 24-29, 35, 37, 49, 51-52] and a highly-dense compact coating with complete coverage conforming to electrical-insulation requirements in harsh LM environment of interest. After exposure of over 520 h in PbLi, thermal cycling within PbLi environment was conducted to establish insulation integrity during possible temperature gradients in a LM circuit. Considering the normal operational temperature of PbLi circulations facilities, 300°C was taken as the base temperature for test campaign-1 while IR at the start of last thermal cycle was taken as the base IR value. To gain higher accuracy in measurements, a digital IR tester (Make: Fluke, Model: 1550C) with ±5% uncertainty was used. Measured IR values are reported under **Table-3** and IR derating trends are shown in **Fig.5**.

In addition to temperature effect, observed IR derating trends also include the effect of rising PbLi level surrounding the coated probes. However, for the current set-up, overall level rise per 10°C change is estimated less than 1 mm. In consideration to the chemical stability of $Al_2O_3$, temperature is assumed to be the prominent derating factor for a conservative estimation of IR derating. The above assumption is supported by the rising trend of IR values towards the base IR value as mentioned in the last row of **Table-3**, taken just after achieving 300°C at the surface of tank, corroborating thermal and chemical stability of $Al_2O_3$ insulation in high-temperature PbLi environment. It should be noted that these derating factors have been estimated post-exposure of the insulation to high-temperature corrosive PbLi environment for over 690 h. In view of similarity of the fabricated probes with electrical cables, observed temperature-based IR derating is much better [55].

**Fig.6** presents images of P1 and P2 after removal and post-chemical cleaning, depicting only partial cleaning achieved even after sufficient immersion time in a 1:1:1 (volume ratio) solution of acetic acid, hydrogen-peroxide and ethyl alcohol. Such a change in appearance was not observed during an earlier short-duration experimental study, where complete cleaning could be achieved for one of the sample probes exposed to PbLi at 300°C for continuous 250 h. Observed discoloration in present case could be attributed to partial ingress of PbLi in the coated layer over longer exposure durations. After chemical cleaning, no weight loss was observed for P1 while a normalized weight loss of ~ 1.56 mg/cm² over the coated surface area was observed for P2. This loss could be primarily accounted for dehydration of topmost untreated coat layer of P2 during facility heating and could also include partial thinning of the coating section immersed in PbLi. However, in view of no variation in the order of magnitude for IR within uncertainty limits at relevant temperatures (refer **Table-2** and **Table-3),** dehydration could be ascribed as the prominent cause for observed weight loss.

### 3.2.2. Test campaign-2

Electrical-insulation performance achieved for probe P2 (test campaign-1) through gradual heat-curing of uppermost layer during system operation seems promising to achieve coating on substrates with sharp bends, multiple/parallel internal flow sections, complex geometries etc. To establish the repeatability and reliability of this coating technique, a more rigorous test campaign-2 was conducted using probe P3 as per details shown in **Fig.7**, while the measured IR values with time and temperature are reported under **Table-4**.

The variation in IR followed a pattern similar to that observed during test campaign-1, accounting for gradual heat-curing of the untreated coat layer. After continuous high-temperature PbLi exposure of over 1340 h, temperature based derating factors were estimated using a base temperature of 350°C as reported under **Table-5** and **Fig.8**. For insulation health estimations, IR readings were taken at 60 s and 120 s after application of high voltage for all the cases.

As observed from **Table-5**, at any given temperature, a rising value of IR from 60 s to 120 s substantiates the health of insulation. A comparison of data from **Table-4** and **Table-5**, alongwith consideration to the fact that capacitive charging current normally decays within 30 seconds, signifies that dielectric polarization is taking more time which is also indicative of insulation health [55]. However, as the order of magnitude for IR readings at 60 s and 120 s remains the same, this effect is not of any significant implication for practical purposes of achieving electrical-insulation in a LM environment. **Fig.8** also affirms that temperature deratings estimated using IR data at 120 s are higher than corresponding deratings using IR data at 60 s. However, degradation in the IR over a temperature gradient of 50°C is relatively low compared to both the cases under test campaign-1. Also, the IR value essentially did not deteriorate between 350°C-370°C in contrast to the pattern observed under test campaign-1. These improvements could all be ascribed to long duration exposure near the required curing temperature of 427°C. The IR after cool down to the base temperature was observed higher than the corresponding value at start of the cycle, which is assumed to be the result of an effective dielectric polarization due to repetitive IR tests at high voltage leading to a polarization current component to zero. This assumption is also in coherence to the nearly identical IR values observed at 60 s and 120 s after a few IR measurements (refer **Table-5**). However, this needs to be investigated in more detail with similar tests. The IR at base temperature of 350°C measured at the end of last thermal cycle (1350 h of exposure) tends towards the measured IR observed at same temperature between 500 h - 800 h of exposure. Similarly, the measured IR values at 300°C (1367 h of PbLi exposure) were 1.89 x $10^8$ Ω and 3.32 x $10^8$ Ω after 60 s and 120 s, respectively. All these observations corroborate high integrity and healthiness of electrical-insulation over complete test duration. Image of the probe after removal from the tank and after chemical cleaning is shown in **Fig.9**.

The PbLi exposed coated section appeared blackish after cleaning, which is in close agreement to the findings reported in [56]. **Fig.9** shows major visible cracks just after completion of chemical cleaning. Spontaneous chipping of a few coat-sections was noticed during chemical cleaning followed by detachment of almost complete coating during subsequent drying at RT. Therefore, weight measurements for P3 post-exposure could not be performed. A similar observation with an AIN-BN sample immersed in Li was reported [21], where Li covered surface broke into pieces during subsequent cleaning in water, accounting for induced stresses due to reaction between water and Li. However, in present study, Li activity is negligible owing to the low weight percentage of Li (0.62% to 0.68%). Additionally, neither such observations have been reported for alumina coated samples in previous corrosion studies with PbLi nor observed during test campign-1 for probe P2, which had a similar fabrication method and PbLi immersion process as that of P3. The possibility of crack-generation/initiation during thermal cycling cannot be completely ruled out owing to large difference in CTEs. However, the unexposed top-section of coated probe (~20 mm), which is also expected to experience similar thermal gradients, remained well-adhered to the substrate with no presence of cracks or indication of brittleness. A crack-generation within LM would have led to a sharp decline in the measured IR, which was not observed. Therefore, in view of IR integrity observed from **Table-4** and **Table-5**, it is inferred that the through-crack generation occurred post-removal of the probe from LM. One of the possibilities could be the cracking due to compressive stresses generated over LM exposed portion of coated substrate because of temperature-dependent volumetric contraction of the surface-adhered PbLi during cool-down period. Under present experimental constraints, such compressive stresses are inevitable. In contrast, during the flowing PbLi conditions in a coolant/breeder system of a fusion reactor, LM will be surrounded by ceramic insulating coatings, leading to an absence of such compressive stresses. However, as the coating cracks were not observed during test campaign-1, the contribution of PbLi exposure duration and temperature is of interest for further compatibility studies. In view of similarity of the fabricated probes to electrical cables, volumetric electrical-resistivity was estimated for the enveloping cases using IR relation for an electrical cable. It should be noted that in the present experimental study, a conservative estimation of volumetric electrical-resistivity has been performed taking the tip-portion (blob) as a perfect electrical-insulation in view of its relative higher thickness as compared to the coating thickness along the length. Therefore, complete insulation leakage current is assumed to flow only through thickness of coating along the probe length, leading to an under-estimation of calculated volumetric electrical-resistivity for the probe(s) under reported operating conditions. Calculated volumetric electrical-resistivity remained of the order of $10^9$ – $10^{11}$ $\Omega$-cm between 300°C-400°C. The coating resistance, measured as a product of volumetric electrical-resistivity and insulation thickness, remained of the order of $10^4$ – $10^6$ $\Omega$-m$^2$, many orders of magnitude higher than required for successful operation of LM based coolant circuits with acceptable MHD pressure drops [2, 24, 57]. Calculations for volumetric electrical-resistivity and coating resistance are provided under **Appendix-A**. This further substantiates the applicability of $Al_2O_3$ coatings for high-temperature PbLi coolant/breeder circuits in fusion power plants. It can be observed that the volumetric electrical-resistivity for fully treated probe P1 is one order of magnitude higher than that of P2 at same temperature, an effect that can be attributed to higher degree of compaction achieved with full heat-curing. However, sufficient volumetric electrical-resistivity achieved and ease of coat-formations using gradual heat-curing through system operations makes the second method more promising and attractive.

## 4. Metallographic investigations

Dip-coating technique usually results in an uneven coating thickness across the length. Additionally, heat-cure of coated probes in a horizontal orientation, as mentioned under Section-2, may also contribute to coating thickness variations under influence of gravity. In this view, cross-sectional samples were cut across PbLi exposed portions of probes P1 and P2, using a diamond wire-cutter, to examine coating thicknesses and PbLi ingress depth. A representative SEM image for one of the cross-sectional samples from P1, as shown in **Fig.10**, confirms a compact coating free from cracks (cracks appearing on the surface of cross-sections are due to wire-cutting). For a given probe, coating thicknesses were measured at various points of the cut samples and an average coating thickness value was calculated by taking average of all the thickness measurements performed. For illustration purposes, coating estimations on one of the samples cut from probe P2 are depicted in **Fig.11**. Estimations as per method described above resulted in average coating thicknesses of ~495 µm and ~212 µm for PbLi immersed portions of probe P1 and P2, respectively. To estimate PbLi ingress depth, EDX point-analysis was carried out at different locations across the cross-sections. Points indicated in **Fig.12(a)** and **Fig.12(b),** moving radially inwards towards the substrate, were chosen for point EDX analysis of probes P1 and P2, as detailed in **Table-6** (point $P_1$ lying at the surface of the coating exposed to PbLi). The points of analyses for P1 and P2 were ~20 µm apart and ~10 µm apart, respectively. In view of the the primary objective of validating electrical-insulating coating towards application in PbLi eutectic alloy, ingress depth of Pb (~99.32 - 99.38% by weight in PbLi eutectic) was considered representative in consistency with previously reported coating qualification studies [27, 33, 36, 49, 51]. Analyses at various points corroborate Pb ingress limited to 20 µm depth for both the probes tested under test campaign-1. The limited penetration depth of Pb validates $Al_2O_3$ coating as an electrical-insulator as well as corrosion-barrier for high-temperature PbLi applications. Restricted ingress depth also substantiates the assumption of temperature being the prominent factor in IR derating, as discussed under Section-3.2.1.

As shown in **Fig. 13(a)**, the coating thickness for P3, as estimated from the chipped-off coat sections and averaged over 04 data-points, was ~429 µm. A representative image for one of the samples with observed cracks acorss the thickness is shown in **Fig.13(b)**. PbLi ingress depth estimations using EDX were rendered inconclusive as Pb ingress was detected over the complete coating thickness in complete contrast to observed IR reported under **Table-4** and **Table-5**. This can be explained in view of observed major cracks after chemical cleaning (refer **Fig.9** and **Fig. 13(b)**), which seemed to have allowed ingress of Pb from the cleaning solution itself acorss complete coating thickness. All the analysed samples from P3 indicated similar results.

Surface microstructure of PbLi exposed coating portion for P1 is presented in **Fig.14 (a)**, suggesting presence of flaked/whisker-type structures (marked as A) and agglomerates less than 1 µm in size (marked as B) mostly covered at the surface with flakes/whiskers. The typical flake/whisker-type pattern is in close agreement to the reportings of $\alpha$-$Al_2O_3$ in [58-59] while the globular pattern is generally accosiated with $\theta$-phase of $Al_2O_3$ [59]. X-ray Diffraction (XRD) analysis of the sample, as shown in **Fig.14 (b)**, depicts the presence of only $\alpha$-phase of $Al_2O_3$ (ICDD card no: 02-002-1227) . Prominent presence of $\alpha$-phase is attributed to the calcined $Al_2O_3$ powder utilized as a raw ingredient for coating suspension, as verified through separate XRD analysis of dried coating suspension sample. As reported under various studies, $\alpha$-$Al_2O_3$ is thermodynamically stable phase formed through irreversible phase-transformations from other transition aluminas when heated at temperature in excess of 1000ºC, resulting in the crystal shapes varying with the nature of precursor [59]. Therefore, the $\alpha$-$Al_2O_3$ agglomerates seem to have been converted from θ-Al$_2$O$_3$ particles coalesced in larger globules. Aluminium monophosphate (AlPO$_4$) detected in the XRD (ICDD card no: 00-048-0652) is the binding agent of utilized coating while the presence of lead-oxides (PbO, PbO$_2$) is primarily accounted for oxidation of Pb content adhered to the probe coated surface and pre-existing oxides in used PbLi melt.

Formation of Li$_2$Al$_2$O$_4$ (ICDD card no: 00-001-1306), instead of generally observed Li$_2$O for PbLi melt systems, is in agreement to previously reported compatibiity studies between Al$_2$O$_3$ and PbLi [36, 49, 60]. Previous investigations affirm that Li reacts with trace amount of oxygen present in the argon gas to produce Li$_2$O, which further reacts with Al$_2$O$_3$ over time to form LiAlO$_2$ . Regular filling/purging of argon cover gas to maintain an inert atmosphere over liquid-metal acts as a constant supply souce for oxygen in the present experimental study. The particle morphology of the PbLi exposed section and the corresponding XRD analysis for P3, as shown in **Fig.15(a)** and **Fig. 15(b)** respectively, also confirm the presence of α-Al$_2$O$_3$ with nearly spherical and uniform particle size < 100 nm.

An important establishment from the XRD results for all the tested probes is the prominent presence of α-Al$_2$O$_3$ and AlPO$_4$, both of which are promising candidates for TPB with high permeation reduction factors and exhibit high mechanical and chemical stability at relevant operating temperatures [61-67]. This observation alongwith successful coating trials conducted on different substrate materials including P-91 sheets and internals of SS-316L pipe-sections paves the way towards exhaustive qualification of the studied coating technique for nuclear fusion applications.

## Conclusions

An aluminium monophosphate-bonded high purity Al$_2$O$_3$ coating suspension applied using dip-coating technique was studied as a candidate functional material towards electrical-insulation requirements in liquid Pb-16Li breeder applications. The coating method required relatively low-temperature exposure of the substrate and demonstrated good adherence, compactness, non-porosity and high insulation resistance at room-temperature. Coated substrates were rigorously validated in static PbLi for continuous 1360 h in the temperature range of 300ºC-400ºC for assessment of electrical-insulation integrity and temperature derating. Al$_2$O$_3$ coated substrates demonstrated excellent chemical compatibility, thermal stability and electrical-insulation characteristics with estimated coating resistance of the order of 10$^4$ – 10$^6$ Ω-m$^2$, better than required towards practical applications in MHD pressure drop reductions and successful realization of PbLi liquid-metal based breeders. This leaves an ample margin for insulation degradation over long duration operations and through radiation induced conductivity. EDX analysis corroborates Pb ingress limited to within 20 µm of coating thickness for temperature upto 350ºC over time duration of 700 h. For its ease of application, the adopted coating method seems compatible for coating of difficult substrate geometries like manifolds and complex structures typical of breeder blankets. High brittleness observed after chemical-cleaning of PbLi exposed coated-section is of interest towards further studies. Prominent presence of thermodynamically stable α-$Al_2O_3$ further advocates the coating utility towards tritium permeation barrier applications in nuclear fusion power plants. Homogeneity of the coating application method was established for different substrate materials and geometries including internals of pipe-sections. However, a complete corroboration towards application suitability requires further validation under flowing PbLi conditions at representative temperature and radiation environment to assess extent of radiation-induced conductivity.

## Acknowledgements

Authors are grateful to Mr. A. Zala and Dr. N. I. Jamanapara (FCIPT, IPR) towards $Al_2O_3$ coating trials on SS-316L samples using plasma assisted tempering and thermal treatments, Mr. A. Satyaprasad (FCIPT, IPR) for diamond wire-cutting of coated samples and Mr. H. Tailor (IPR) for support during operation of test facility. Authors also deeply appreciate the invaluable discussions with Mr. V. Mahesh and Mr. V. Ranjan (IPR) towards the experimental methodology.

## Appendix-A: Estimation of volumetric electrical-resistivity and coating resistance

Insulation resistance of an electrical-cable is given by the following relation:

$$R_i = \left(\frac{\varrho}{2\pi l}\right) \ln \frac{R}{r} \quad \dots\dots\dots\dots\dots\dots\dots\dots\dots\text{Eq. (A.1)}$$

Where, $R_i$ = Insulation resistance of the cable ($\Omega$), $\varrho$ = Volumetric electrical-resistivity of insulation material ($\Omega$-m), l = Length of the insulation over the cable (m), R = Outer radius of the cable with insulation (m) and r = Outer radius of the conductor (m).

Using data from Table-1, 2, 3, 4 & 5 and taking the enveloping IR values at each temperature:

| Test Campaign | Probe | Insulation Resistance ($\Omega$) | Average coating thickness (mm) | Coated length (m) | Temperature (ºC) | Resistivity ($\Omega$-cm) | Coating resistance ($\Omega$-m$^2$) |
|---|---|---|---|---|---|---|---|
| 1 | P1 | $6 \times 10^9$ | 0.495 | 0.085 | 300 | $6.65 \times 10^{11}$ | $3.29 \times 10^6$ |
| | | $6.18 \times 10^9$ | 0.495 | 0.085 | 300 | $6.85 \times 10^{11}$ | $3.39 \times 10^6$ |
| | | $0.822 \times 10^9$ | 0.495 | 0.085 | 350 | $9.12 \times 10^{10}$ | $4.51 \times 10^5$ |
| | | $1 \times 10^9$ | 0.495 | 0.085 | 350 | $1.11 \times 10^{11}$ | $5.49 \times 10^5$ |
| | P2 | $1 \times 10^8$ | 0.212 | 0.065 | 300 | $1.74 \times 10^{10}$ | $3.68 \times 10^4$ |
| | | $3.15 \times 10^8$ | 0.212 | 0.065 | 300 | $5.47 \times 10^{10}$ | $1.16 \times 10^5$ |
| | | $0.835 \times 10^8$ | 0.212 | 0.065 | 350 | $1.45 \times 10^{10}$ | $3.08 \times 10^4$ |
| | | $1 \times 10^8$ | 0.212 | 0.065 | 350 | $1.74 \times 10^{10}$ | $3.68 \times 10^4$ |
| 2 | P3 | $2.90 \times 10^8$ | 0.429 | 0.060 | 300 | $2.55 \times 10^{10}$ | $1.09 \times 10^5$ |
| | | $4.57 \times 10^8$ | 0.429 | 0.060 | 300 | $4.01 \times 10^{10}$ | $1.72 \times 10^5$ |
| | | $1.29 \times 10^8$ | 0.429 | 0.060 | 350 | $1.13 \times 10^{10}$ | $4.86 \times 10^4$ |
| | | $2.99 \times 10^8$ | 0.429 | 0.060 | 350 | $2.63 \times 10^{10}$ | $1.13 \times 10^5$ |
| | | $0.583 \times 10^8$ | 0.429 | 0.060 | 400 | $5.12 \times 10^9$ | $2.20 \times 10^4$ |
| | | $0.768 \times 10^8$ | 0.429 | 0.060 | 400 | $6.74 \times 10^9$ | $2.89 \times 10^4$ |

# References

1. D. Giulio, D. Suarez, L. Batet et al., Analysis of flow channel insert deformations influence on the liquid metal flow in DCLL blanket channels, Fusion Engineering and Design 157 (2020) 111639.

2. M. González and M. Kordac, Electrical resistivity behaviour of alumina flow channel inserts in PbLi, Fusion Engineering and Design 159 (2020) 111761.

3. A. Tassone, G. Caruso and A. D. Nevo, Influence of PbLi hydraulic path and integration layout on MHD pressure losses, Fusion Engineering and Design 155 (2020) 111517.

4. D. Iadicicco, M. Vanazzia, F.G. Ferré et al., Multifunctional nanoceramic coatings for future generation nuclear systems, Fusion Engineering and Design 146, Part B (2019) 1628-1632.

5. L. Bühler and C. Mistrangelo, Pressure drop and velocity changes in MHD pipe flows due to a local interruption of the insulation, Fusion Engineering and Design 127 (2018) 185-191.

6. I. Fernández-Berceruelo, M. Gonzalez, I. Palermo et al., Large-scale behavior of sandwich-like FCI components within the EU-DCLL operational conditions, Fusion Engineering and Design 136, Part A (2018) 633-638.

7. A.F. Rowcliffe, L.M. Garrison, Y. Yamamoto et al., Materials challenges for the fusion nuclear science facility, Fusion Engineering and Design 135 (2018) 290–301.

8. D. Rapisarda, I. Fernandez, I. Palermo et al., Status of the engineering activities carried out on the European DCLL, Fusion Engineering and Design 124 (2017) 876–881.

9. M. Abdou, N. B. Morley, S. Smolentsev et al., Blanket/first wall challenges and required R&D on the pathway to DEMO, Fusion Engineering and Design, 100 (2015) 2-43.

10. S. Smolentsev, N. B. Morley, M. Abdou and S. Malang, Dual-coolant lead–lithium (DCLL) blanket status and R&D needs, Fusion Engineering and Design 100 (2015) 44–54.

11. K. Singh, A. Fernandes, B. Paul et al., Preparation and investigation of aluminized coating and subsequent heat treatment on 9Cr–1Mo Grade 91 steel, Fusion Engineering and Design 89 (11) (2014) 2534-2544.

12. T. Tanaka and T. Muroga, Oxide coating fabrication by metal organic decomposition method for liquid blanket systems, Fusion Engineering and Design 88, 9–10 (2013) 2569-2572.

13. S. Malang, A.R. Raffray and N.B. Morley, An example pathway to a fusion power plant system based on lead–lithium breeder: Comparison of the dual-coolant lead–lithium (DCLL) blanket with the helium-cooled lead–lithium (HCLL) concept as initial step, Fusion Engineering and Design, 84 (2009) 2145-2157.

14. T. Shikama, R. Knitter, J. Konys et al., Status of development of functional materials with perspective on beyond-ITER, Fusion Engineering and Design 83 (2008) 976–982.

15. E. R. Kumar, C. Danani, I. Sandeep et al., Preliminary design of Indian Test Blanket Module for ITER, Fusion Engineering and Design 83 (2008) 1169–1172.

16. B.A. Pint, J.L. Moser and P.F. Tortorelli, Liquid metal compatibility issues for test blanket modules, Fusion Engineering and Design 81 (2006) 901–908.

17. S. Smolentsev, N. Morley and M. Abdou, Code development for analysis of MHD pressure drop reduction in a liquid metal blanket using insulation technique based on a fully developed flow model, Fusion Engineering and Design 73 (1) (2005) 83-93.

18. M. Nakamichi and H. Kawamura, Out-of-pile characterization of $Al_2O_3$ coating as electrical insulator, Fusion Engineering and Design, 58–59 (2001) 719-723.

19. T. Terai, T. Mitsuyama and T. Yoneoka, Fabrication and properties of ceramic coatings for CTR liquid blanket by sputtering method, Fusion Engineering and Design, 51–52 (2000) 207–212.

20. S. Tanaka, Y. Ohara and H. Kawamura, Blanket R&D activities in Japan towards fusion power reactors, Fusion Engineering and Design 51–52 (2000) 299–307.

21. T. Mitsuyama, T. Terai, T. Yoneoka and S. Tanaka, Compatibility of insulating ceramic materials with liquid breeders, Fusion Engineering and Design 39–40 (1998) 811–817.

22. K. Natesan, C.B. Reed and R.F. Mattas, Assessment of alkali metal coolants for the ITER blanket, Fusion Engineering and Design, 27 (1995) 457-466.

23. L. Giancarli, G. Benamati, S. Malang et al., Overview of EU activities on DEMO liquid metal breeder blankets, Fusion Engineering and Design 27 (1995) 337-352.

24. S. Malang, H.U. Borgstedt, E.H. Farnum et al., Development of insulating coatings for liquid metal blankets, Fusion Engineering and Design 27 (1995) 570-586.

25. S. E. Wulf, N. Holstein, W. Krauss and J. Konys, Influence of deposition conditions on the microstructure of Al-based coatings for applications as corrosion and anti-permeation barrier, Fusion Engineering and Design 88 (2013) 2530– 2534.

26. G.K. Zhang, C.A. Chen, D.L. Luo and X.L. Wang, An advance process of aluminum rich coating as tritium permeation barrier on 321 steel workpiece, Fusion Engineering and Design 87 (2012) 1370– 1375.

27. Bhaskar Paul, K. Raju, M. Vadsola et al., Investigations on wear and liquid metal corrosion behavior of aluminized IN-RAFMS, Fusion Engineering and Design 128 (2018) 204–214.

28. S. Feng, Y. Wang, C. Zhang et al., Preparation of $Al_2O_3/Cr_2O_3$ tritium permeation barrier with combination of pack cementation and sol–gel methods, Fusion Engineering and Design 131 (2018) 1–7.

29. C. Zhu, W. Zhang, L. Wang et al., Effect of thermal cycles on structure and deuterium permeation of $Al_2O_3$ coating prepared by MOD method, Fusion Engineering and Design 159 (2020) 111750.

30. T. Chikada, A. Suzuki, Z. Yao et al., Deuterium permeation behavior of erbium oxide coating on austenitic, ferritic, and ferritic/martensitic steels, Fusion Engineering and Design 84 (2009) 590– 592.

31. L.V. Boccaccini, G. Aiello, J. Aubert et al., Objectives and status of EUROfusion DEMO blanket studies, Fusion Engineering and Design 109–111 (2016) 1199–1206.

32. N. Casal, A. García, A. Ibarra et al., Tritium permeation experiment at IFMIF Medium Flux Test Module, Fusion Engineering and Design 84 (2009) 559–564.

33. M. C. Gazquez, S. Bassini, T. Hernandez and M. Utili, $Al_2O_3$ coating as barrier against corrosion in Pb-17Li, Fusion Engineering and Design 124 (2017) 837-840.

34. M. Utili, S. Bassini, L. Boccaccini et al., Status of Pb-16Li technologies for European DEMO fusion reactor, Fusion Engineering and Design 146 (2019) 2676–2681.

35. J. Purushothaman, R. Ramaseshan, S. K. Albert et al., Influence of surface roughness and melt superheat on HDA process to form a tritium permeation barrier on RAFM steel, Fusion Engineering and Design 101 (2015) 154-164.

36. U. Jain, A. Mukherjee, S. Sonak et al., Interaction of alumina with liquid Pb83Li17 alloy, Fusion Engineering and Design 89 (11) (2014) 2554-2558.

37. A. B. Zala, N. I. Jamnapara, C. S. Sasmal et al., Investigation of alumina film formed over aluminized RAFM steel by plasma assisted heat treatment, Fusion Engineering and Design 146-Part B (2019) 2002-2006.

38. Handbook on Lead-bismuth Eutectic Alloy and Lead Properties, Materials Compatibility, Thermal hydraulics and Technologies, NEA, 2015 Edition.

39. Y. Saito, K. Mishima, Y. Tobita et al., Measurements of liquid–metal two-phase flow by using neutron radiography and electrical conductivity probe, Experimental Thermal and Fluid Science 29 (2005) 323–330.

40. J. L. Muñoz-Cobo, S. Chiva, S. Méndez et al., Development of Conductivity Sensors for Multi-Phase Flow Local Measurements at the Polytechnic University of Valencia (UPV) and University Jaume I of Castellon (UJI), Sensors 17(5) (2017)1077.

41. Y. Saito, Bubble Measurements In Liquid-Metal Two-Phase Flow By Using A Four-Sensor Probe, Multiphase Science and Technology, 24 (4) (2012) 279-297.

42. P. Gherson and P. S. Lykoudis, Local measurements in two-phase liquid-metal magneto-fluid-mechanic flow, Journal of Fluid Mechanics 147 (1984) 81-104.

43. C. Courtessole, S. Smolentsev, T. Sketchley and M. Abdou, MHD PbLi experiments in MaPLE loop at UCLA, Fusion Engineering and Design 109-111 (2016) 1016-1021.

44. S. Tosti, A. Moriani and A. Santucci, Design and manufacture of an oven for high temperature experiments of erosion–corrosion of SiC$_f$/SiC into LiPb, Fusion Engineering and Design 88 (2013) 2479-2483.

45. M. Kumar, A. Patel, A. Jaiswal et al., Engineering design and development of lead lithium loop for thermo-fluid MHD studies, Fusion Engineering and Design 138 (2019) 1-5.

46. S. S. Atchutuni, A. Saraswat, C. S. Sasmal et al., Corrosion experiments on IN-RAFM steel in flowing lead-lithium for Indian LLCB TBM, Fusion Engineering and Design 132 (2018) 52-59.

47. A. Saraswat, A. Prajapati, R. Bhattacharyay et al., Experimental investigations on bubble detection in water-air two-phase vertical columns, Chapter-48, Recent Advances in Mechanical

Engineering, Series Title: Lecture Notes in Mechanical Engineering, 2021, Edition:1 (Article in press).

48. http://www.matweb.com/index.aspx

49. N. I. Jamnapara, A. S. Sree, E. R. Kumar et al., Compatibility study of plasma grown alumina coating with Pb–17Li under static conditions, Journal of Nuclear Materials 455 (2014) 612–617.

50. E. Vassallo, M. Pedroni and V. Spampinato, Effect of alumina coatings on corrosion protection of steels in molten lead, Journal of Vacuum Science & Technology B 36 (2018) 01A105.

51. S. E. Wulf, W. Krauss and J. Konys, Long-term corrosion behavior of Al-based coatings in flowing Pb–15.7Li, produced by electrochemical ECX process, Nuclear Materials and Energy 16 (2018) 158-162.

52. A. S. Sree and E. R. Kumar, Effect of Heat Treatment and Silicon Concentration on Microstructure and Formation of Intermetallic Phases on Hot Dip Aluminized Coating on Indian RAFMS, Fusion Science and Technology 65(2) (2014) 282-291.

53. B. Schulz, Thermophysical properties of the Li(17) Pb(83) alloy, Fusion Engineering and Design 14 (1991) 199-205.

54. D. Martelli, A. Venturini and M. Utili, Literature review of lead-lithium thermophysical properties, Fusion Engineering and Design 138 (2019) 183-195.

55. https://www.instrumart.com/assets/Megger-Guide-to-Insulation-Testing.pdf

56. H.U. Borgstedt, H. Glasbrenner and Z. Peric, Corrosion of insulating layers on MANET steel in flowing Pb-17Li, Journal of Nuclear Materials 212-215 (1994) 1501-1503.

57. L. Buhler and S. Molokov. Magnetohydrodynamic flows in ducts with insulating coatings, Rep. KfK-5103, Kernforschungszentrum Karlsruhe, 1993.

58. P. Patel, N. I. Jamnapara, A. Zala and S. D. Kahar, Investigation of hot-dip aluminized Ti6Al4V alloy processed by different thermal treatments in an oxidizing atmosphere, Surface and Coatings Technology 385 (2020) 125323.

59. P. S. Santos, H. S. Santos and S.P. Toledo, Standard transition aluminas. Electron microscopy studies, Materials Research 3 (2000) 104-114.

60. B.A. Pint and K.L. More, Transformation of $Al_2O_3$ to $LiAlO_2$ in Pb–17Li at 800°C, Journal of Nuclear Materials 376 (2008) 108–113.

61. L. Wang, J. Yang, C. Liang et al., Preparation and properties of improved $Al_2O_3$ based MOD coatings as tritium permeation barrier, Fusion Engineering and Design 143 (2019) 233-239.

62. F. Jun, D. Min, J. Fanya et al., Preparation and Properties of Alumina Coatings as Tritium Permeation Barrier by Plasma Electrolytic Oxidation, Rare Metal Materials and Engineering 45(2) (2016) 315-320.

63. D. Levchuk, F. Koch, H. Maier and H. Bolt, Deuterium permeation through Eurofer and α-alumina coated Eurofer, Journal of Nuclear Materials 328 (2004) 103–106.

64. R. Yin, L. Hu, J. Tang at el., In-situ oxidation of aluminized stainless-steel to form alumina as tritium permeation barrier coating, Fusion Engineering and Design 163 (2021) 112154.

65. L. Wang, J.J. Yang, Y.J. Feng et al., Preparation and characterization of $Al_2O_3$ coating by MOD method on CLF-1 RAFM steel, Journal of Nuclear Materials 487 (2017) 280-287.

66. K. Zhang and Y. Hatano, Preparation of Mg and Al phosphate coatings on ferritic steel by wet-chemical method as tritium permeation barrier, Fusion Engineering and Design 85 (2010) 1090–1093.

67. D. Chen, L. He and S. Shang, Study on aluminum phosphate binder and related Al2O3-SiC ceramic coating, Materials Science and Engineering A348 (2003) 29-35.

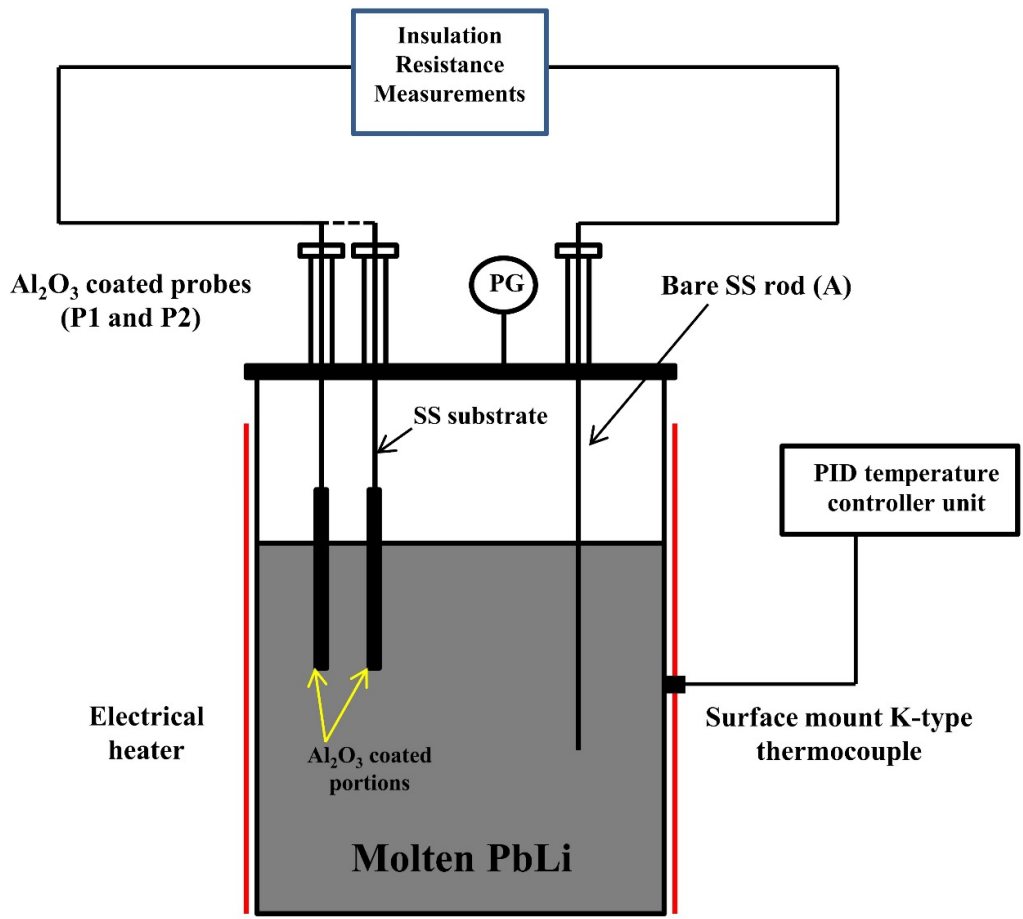

**Fig.1** Insulation resistance measurement scheme for Al₂O₃ coated probes in liquid PbLi

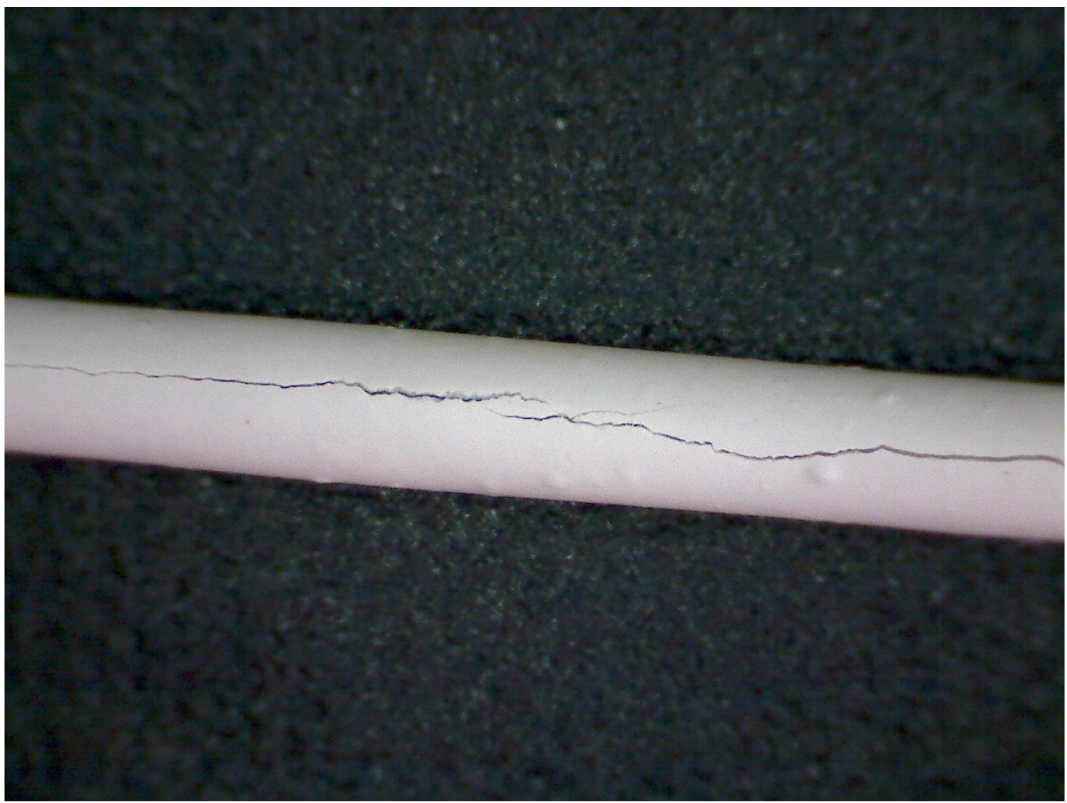

**Fig.2** A full-length crack with blisters on deposited Al₂O₃ coating due to short temperature ramp-up time

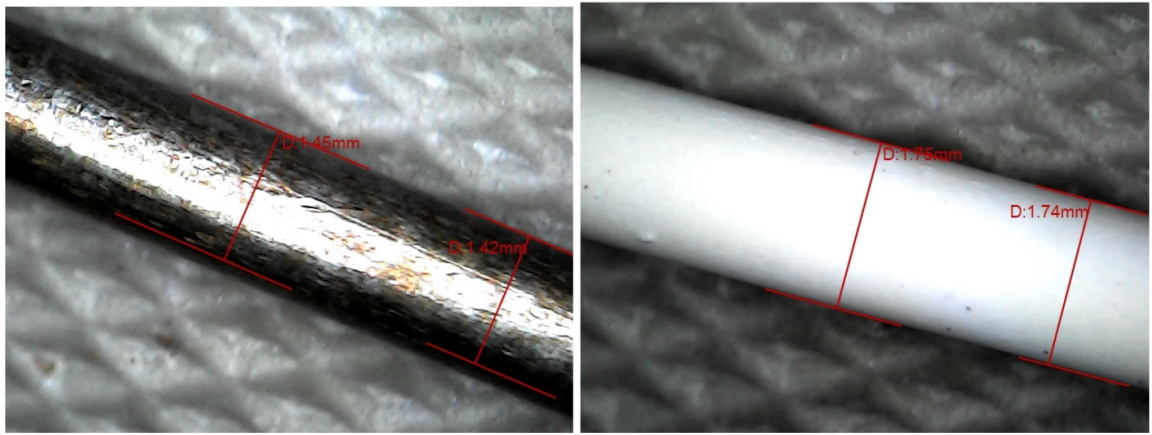

**Fig.3(a)** Al$_2$O$_3$ coated SS-308L electrode (coating thickness ~155 µm)

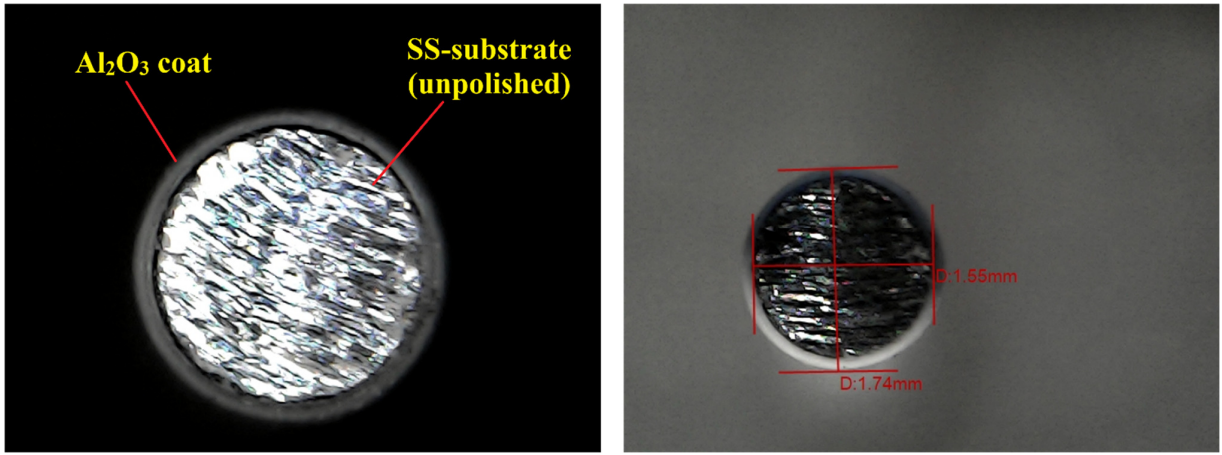

**Fig.3(b)** Al$_2$O$_3$ coated SS-316L electrode with bare SS tip (coating thickness ~95 µm)

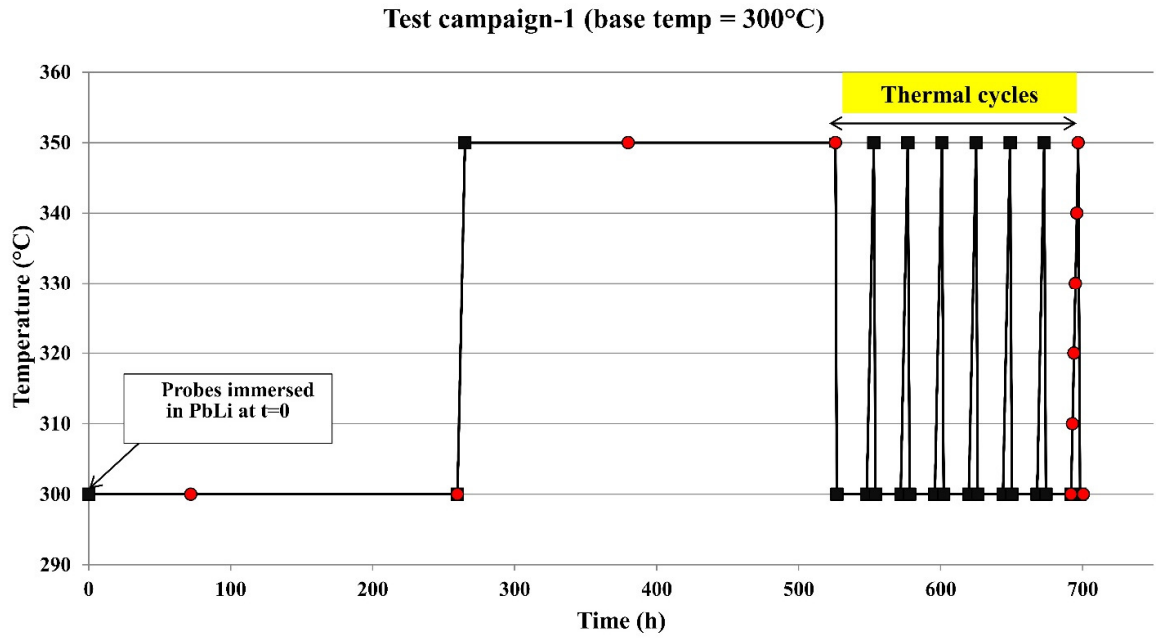

**Fig.4** Temperature and time-duration details for test campaign-1

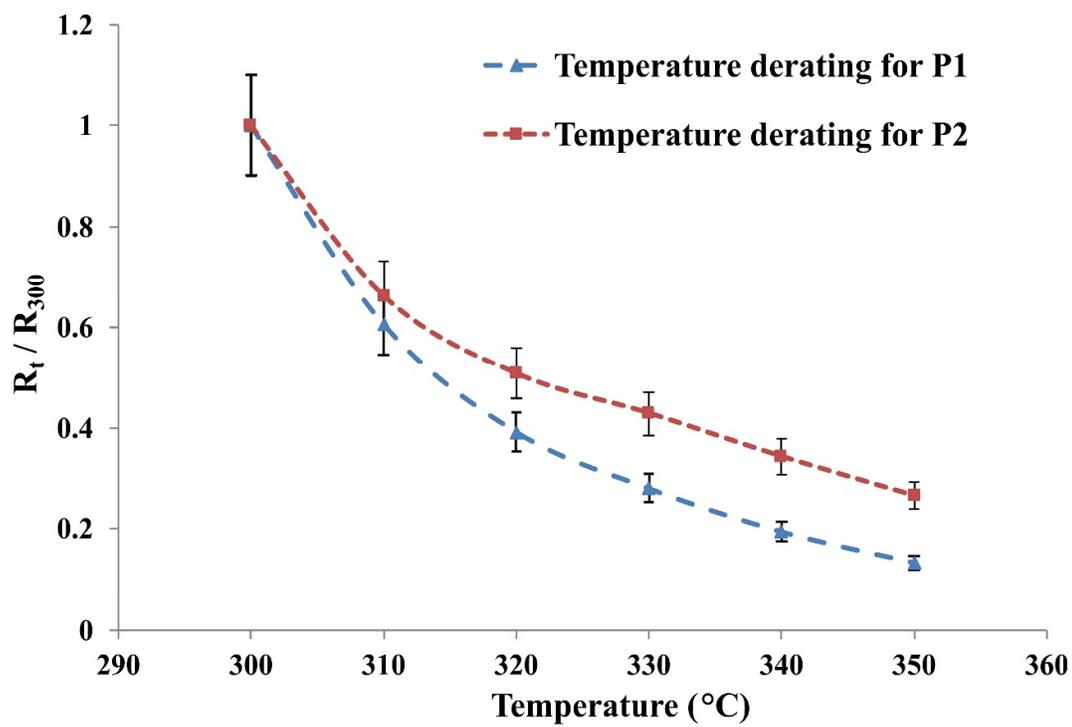

**Fig.5** Temperature related IR derating trends for P1 and P2 (post-exposure to static PbLi for ~690 h)

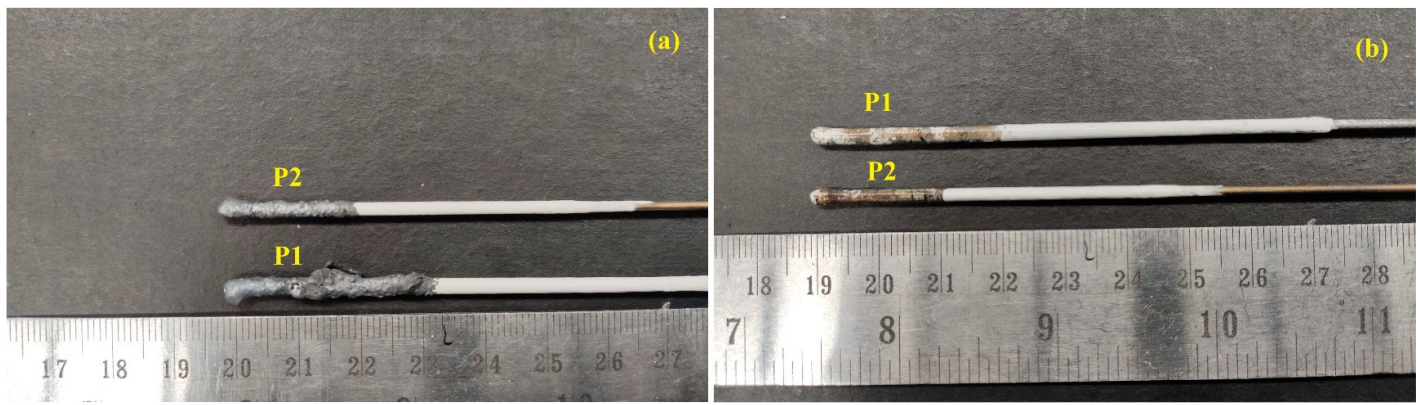

**Fig.6** Condition of P1 and P2 after (a) continuous PbLi exposure of over 700 h; (b) after chemical cleaning

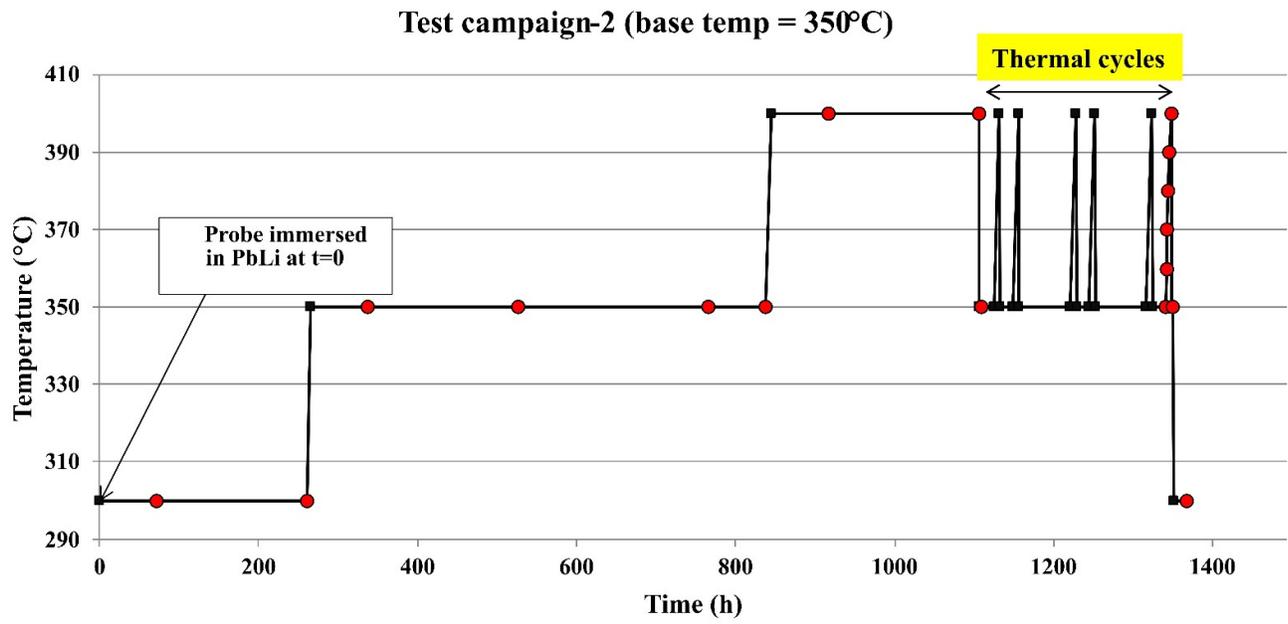

**Fig.7** Temperature and time-duration details for test campaign-2

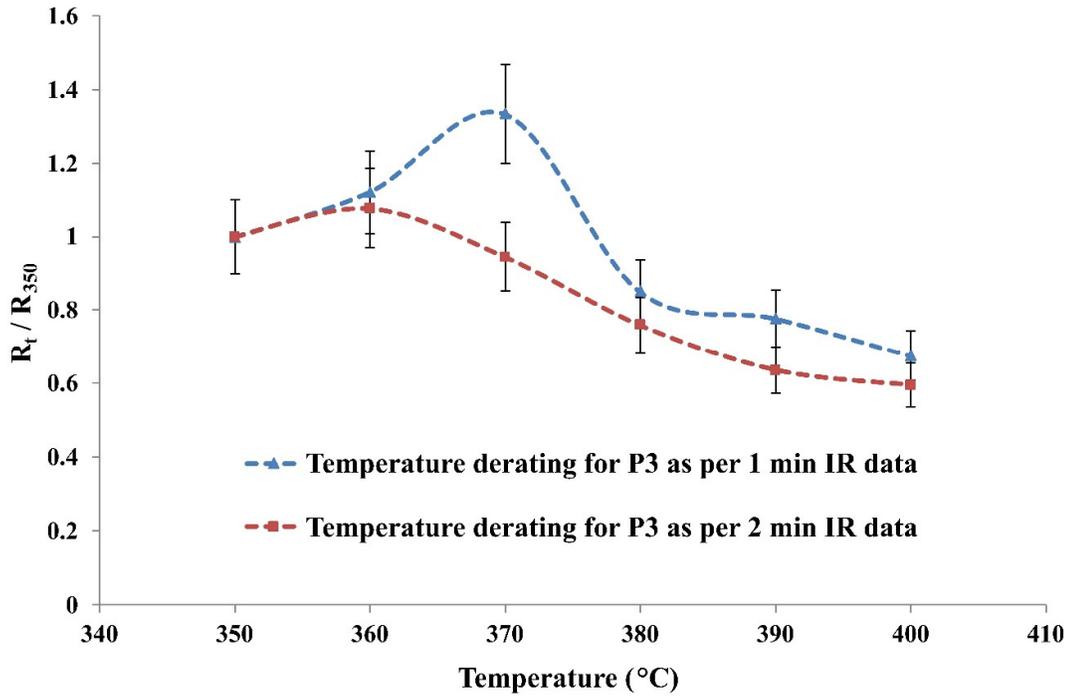

**Fig.8** Temperature related IR derating trend for P3 (post-exposure to static PbLi for ~1340 h)

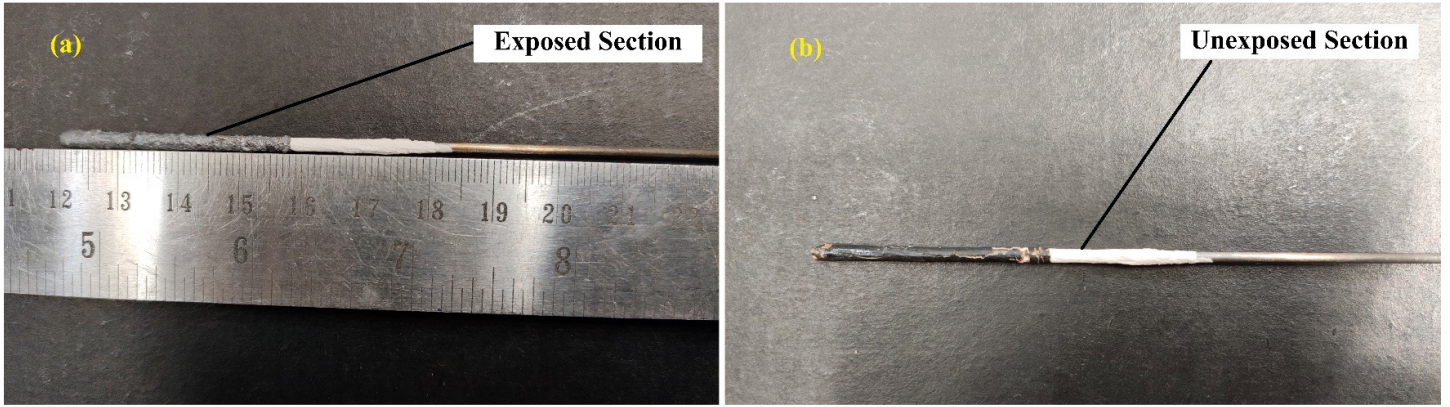

**Fig.9** Condition of probe after (a) continuous PbLi exposure of over 1360 h; (b) after chemical cleaning

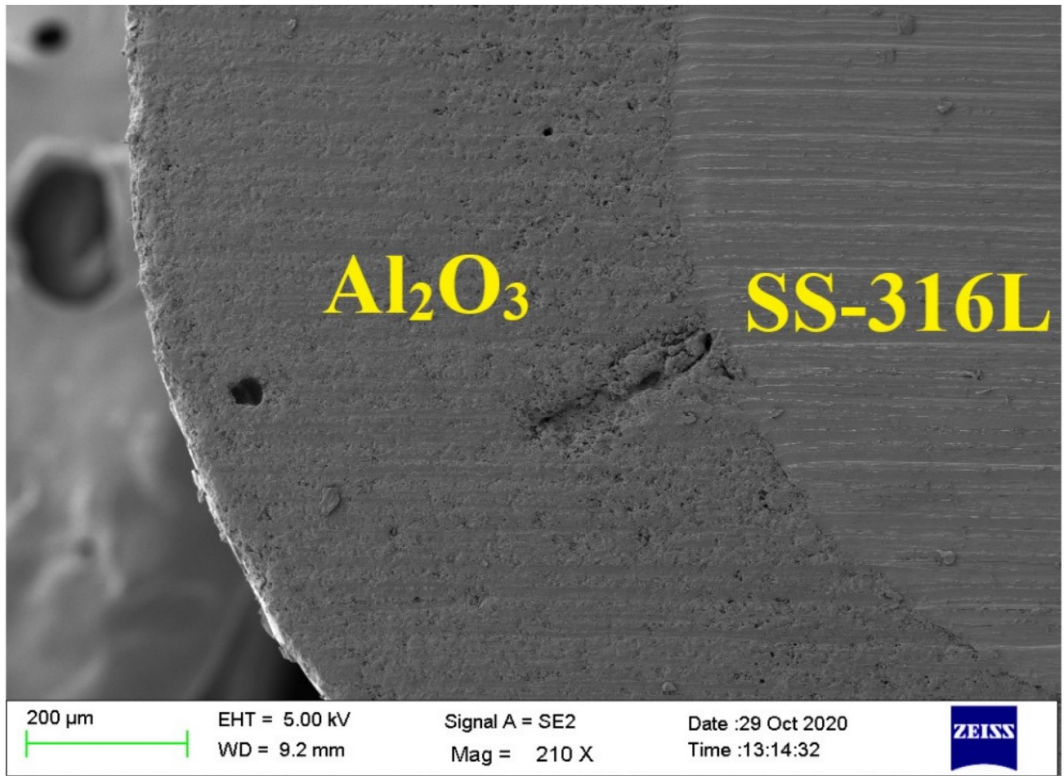

**Fig.10** Cross-sectional view of a sample from P1 showing $Al_2O_3$ coating layer deposited and well adhered to substrate

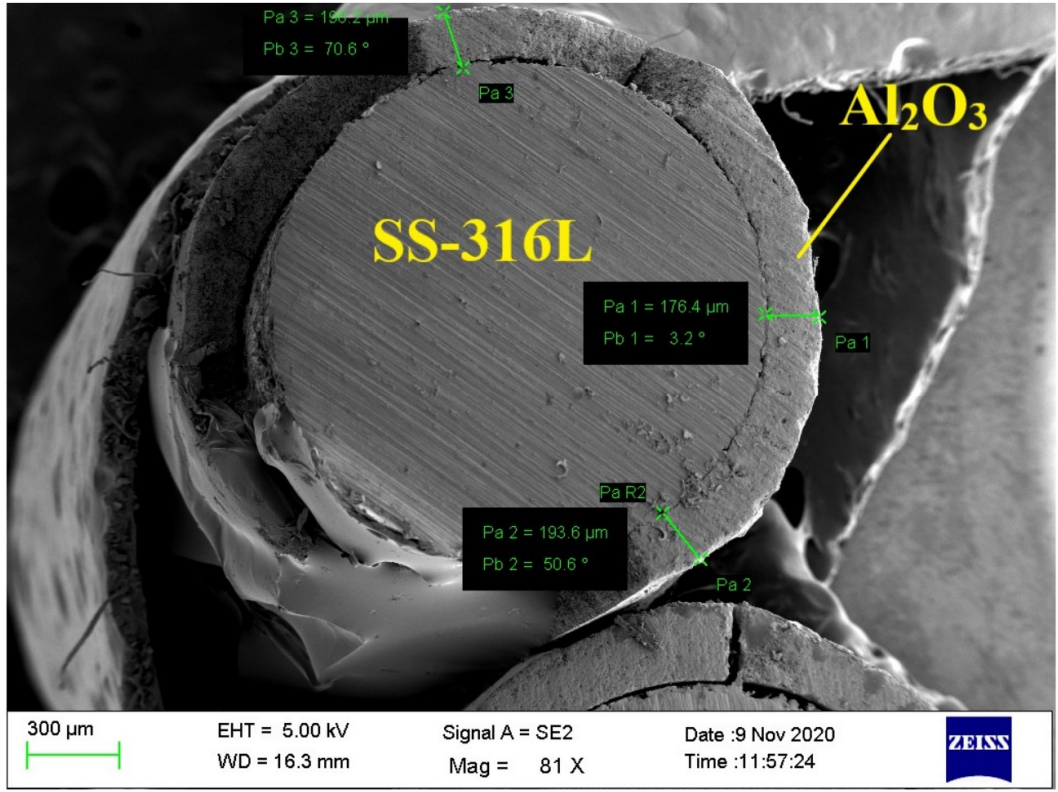

**Fig.11** Cross-sectional view of a sample from P2 for estimation of average coating thickness

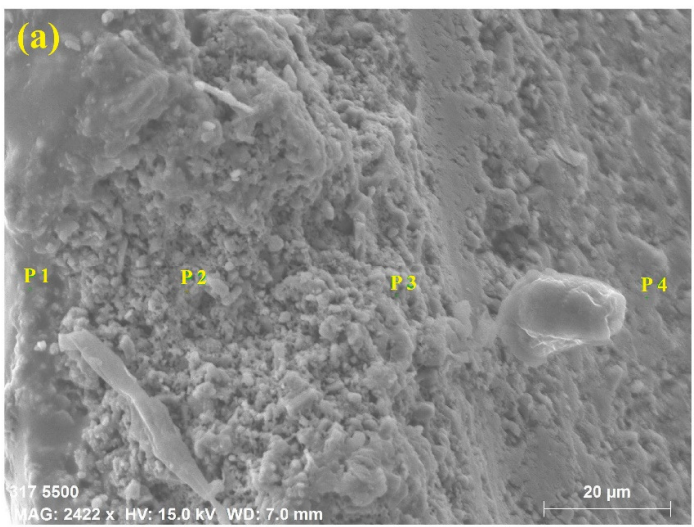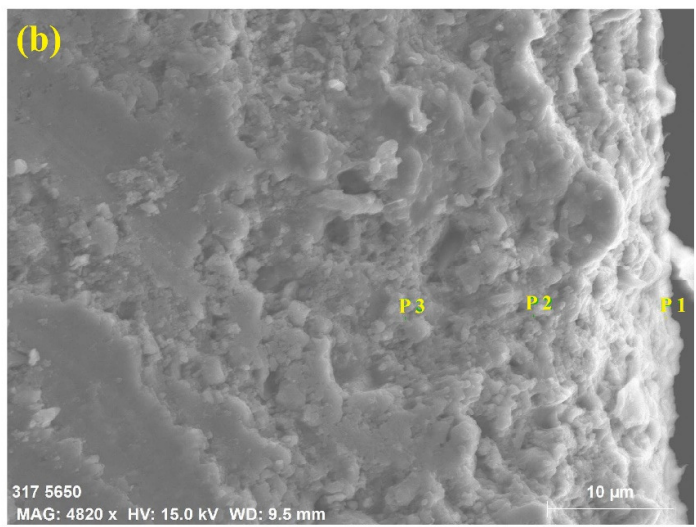

**Fig.12** EDX point-analysis for ingress depth estimations of Pb for (a) Probe P1; (b) Probe P2

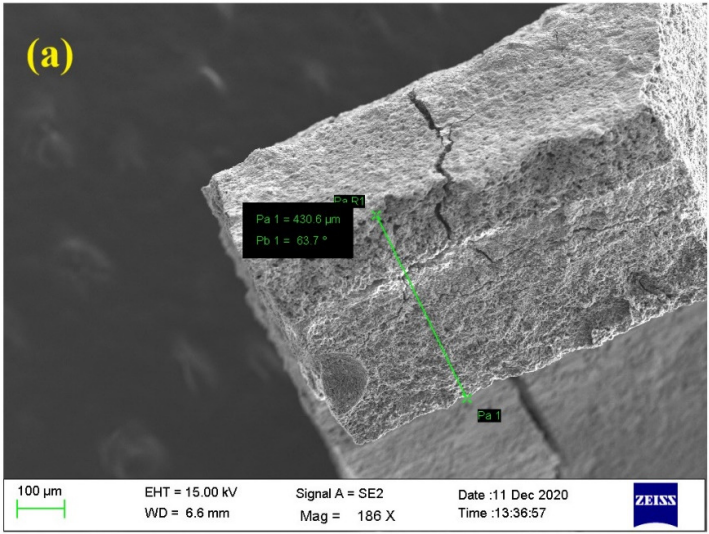 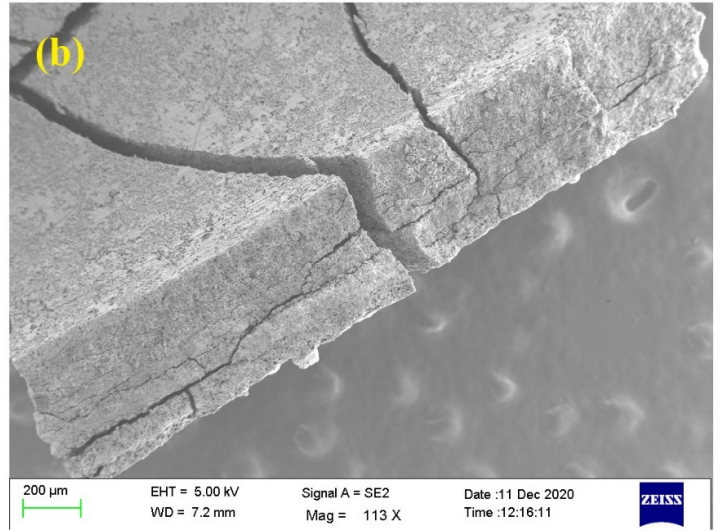

**Fig.13** SEM analysis for Probe P3 (a) Thickness estimations; (b) Cracks across the coating thickness

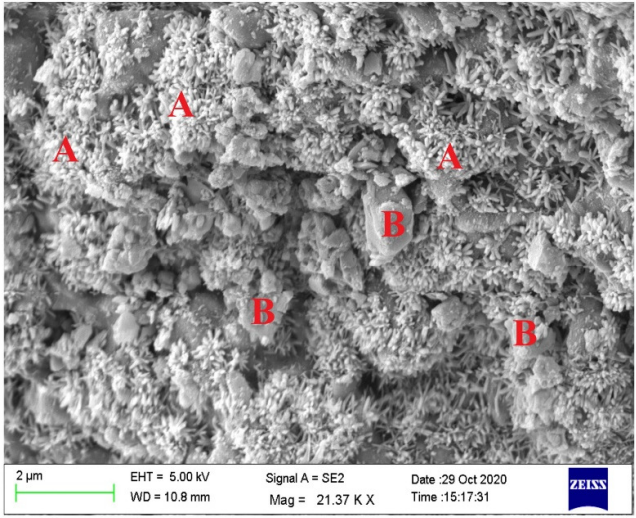

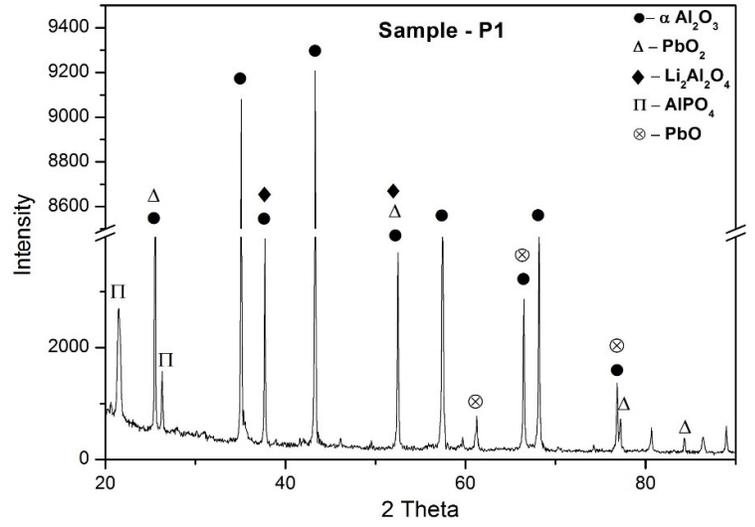

**(a)**

**(b)**

**Fig.14** (a) Surface morphology of PbLi exposed Al$_2$O$_3$ section for P1; (b) XRD pattern

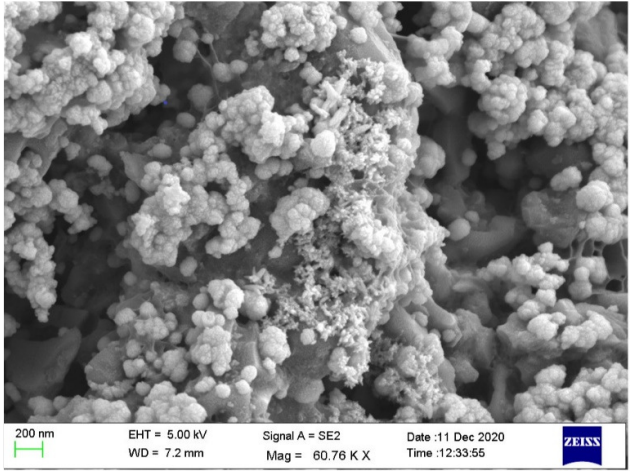

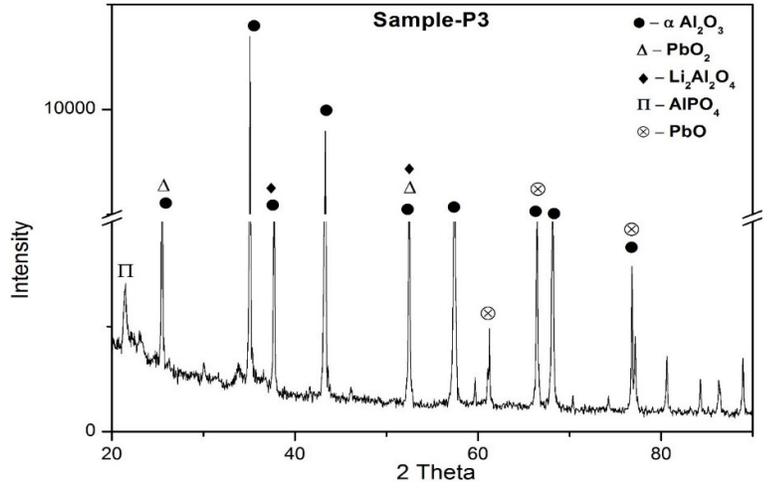

**(a)**

**(b)**

**Fig.15** (a) Detailed particle morphology of PbLi exposed Al$_2$O$_3$ section for P3; (b) XRD pattern

**Table-1**: Details of fabricated probes tested under test campaign-1 and test campaign-2

| Test campaign | Probe identification | Coated length (mm) | Weight (g) |
|---|---|---|---|
| 1 | P1 | ~85 | 8.57521 ± 0.00001 |
| | P2 | ~65 | 7.45867 ± 0.00001 |
| 2 | P3 | ~60 | 7.66220 ± 0.00001 |

**Table-2**: IR measurements for P1 and P2 probes in PbLi environment during test campaign-1

| Time (h) (since t=0) | Temperature (°C) | DC test voltage (V) | IR for Probe P1 (x $10^9$ Ω) | IR for Probe P2 (x $10^8$ Ω) |
|---|---|---|---|---|
| 72 | 300 | 100 | > 6 | > 1 |
| 260 | 300 | 100 | > 6 | > 3 |
| 380 | 350 | 100 | > 1 | > 1 |
| 525 | 350 | 100 | > 1 | > 0.9 |

**Table-3**: Temperature derating factors for IR during last thermal cycle of test campaign-1

| Temperature (°C) | DC test voltage (V) | IR for Probe P1 (x $10^9$ Ω) | Temperature derating for P1 | IR for Probe P2 (x $10^8$ Ω) | Temperature derating for P2 |
|---|---|---|---|---|---|
| 300 (base) | 275 | 6.18 | 1.00 | 3.15 | 1.00 |
| 310 | 275 | 3.75 | 0.61 | 2.09 | 0.66 |
| 320 | 275 | 2.42 | 0.39 | 1.60 | 0.51 |
| 330 | 275 | 1.73 | 0.28 | 1.35 | 0.43 |
| 340 | 275 | 1.20 | 0.19 | 1.08 | 0.34 |
| 350 | 275 | 0.822 | 0.13 | 0.835 | 0.27 |
| 300 | 275 | 6.08 | 0.98 | 2.39 | 0.76 |

**Table-4**: IR measurements for P3 probe in PbLi environment during test campaign-2

| Time (h) (since t=0) | Temperature (°C) | DC test voltage (V) | IR for Probe P3 (x $10^8 \Omega$) |
|---|---|---|---|
| 72 | 300 | 275 | 2.90 |
| 261 | 300 | 275 | 4.57 |
| 338 | 350 | 275 | 1.30 |
| 526 | 350 | 275 | 2.18 |
| 766 | 350 | 275 | 2.99 |
| 838 | 350 | 275 | 2.89 |
| 917 | 400 | 275 | 0.583 |
| 1106 | 400 | 275 | 0.767 |
| 1109 | 350 | 275 | 2.99 |

**Table-5**: Temperature derating factors for IR during last thermal cycle of test campaign-2

| Temperature (°C) | DC test voltage (V) | IR for Probe P3 after 60 s (x $10^8 \Omega$) | Temperature derating for P3 (as per IR data after 60 s) | IR for Probe P3 after 120 s (x $10^8 \Omega$) | Temperature derating for P3 (as per IR data after 120 s) |
|---|---|---|---|---|---|
| 350 (base) | 275 | 0.945 | 1.00 | 1.29 | 1.00 |
| 360 | 275 | 1.06 | 1.12 | 1.39 | 1.08 |
| 370 | 275 | 1.26 | 1.33 | 1.22 | 0.95 |
| 380 | 275 | 0.804 | 0.85 | 0.98 | 0.76 |
| 390 | 275 | 0.735 | 0.78 | 0.821 | 0.64 |
| 400 | 275 | 0.638 | 0.68 | 0.768 | 0.60 |
| 350 | 275 | 2.28 | 2.41 | 2.37 | 1.84 |

**Table-6**: Ingress profile of Pb for probes P1 and P2

| Probe Identification | Point number (marked in yellow) | Normalized Pb % as per EDX analysis |
|---|---|---|
| Probe P1 | $P_1$ (at surface) | 13.08 |
| | $P_2$ (~20 μm from $P_1$) | 0.00 |
| | $P_3$ (~40 μm from $P_1$) | 0.00 |
| Probe P2 | $P_1$ (at surface) | 2.92 |
| | $P_2$ (~10 μm from $P_1$) | 0.69 |
| | $P_3$ (~20 μm from $P_1$) | 0.00 |